\documentclass[superscriptaddress,onecolumn,secnumarabic,nobibnotes,aps,prd,showpacs,nofootinbib,12pt]{revtex4}
\usepackage{eurosym}
\usepackage{graphicx}
\usepackage{epsf}
\usepackage{bm}
\usepackage{amsmath}
\usepackage{amsfonts}
\usepackage{amssymb}
\usepackage{color}
\usepackage{subfig}

\providecommand{\U}[1]{\protect\rule{.1in}{.1in}}

\newcommand{\be}{\begin{equation}}
\newcommand{\ee}{\end{equation}}

\newcommand{\mincir}{\raise
-3.truept\hbox{\rlap{\hbox{$\sim$}}\raise4.truept\hbox{$<$}\ }}
\newcommand{\magcir}{\raise
-3.truept\hbox{\rlap{\hbox{$\sim$}}\raise4.truept\hbox{$>$}\ }}

\begin{document}

\title{Eisenhart-Duval lift, Nonlocal Conservation laws and Painlev\'{e}
Analysis in Scalar field Cosmology}
\author{Andronikos Paliathanasis}
\email{anpaliat@phys.uoa.gr}
\affiliation{Institute of Systems Science \& Department of Mathematics, Faculty of
Applied Sciences, Durban University of Technology, Durban 4000, South Africa}
\affiliation{Centre for Space Research, North-West University, Potchefstroom 2520, South
Africa}
\affiliation{Departamento de Matem\'{a}ticas, Universidad Cat\'{o}lica del Norte, Avda.
Angamos 0610, Casilla 1280 Antofagasta, Chile}

\begin{abstract}
We investigate the existence of nonlocal conservation laws for the
gravitational field equations of scalar field cosmology in an FLRW
background with a dust fluid source. We perform such analysis by using a
novel approach for the Eisenhart-Duval lift. It follows that the scalar
field potential $V\left( \phi \right) =\alpha \left( e^{\lambda \phi }+\beta
\right) $ admits nontrivial conservation laws. Furthermore, we employ the
Painlev\'{e} analysis to examine the integrability of the field equations.
For the quintessence model, we establish that the cosmological field
equations possess the Painlev\'{e} property and are integrable for $\lambda
^{2}>6$. In contrast, for the phantom scalar field, the cosmological field
equations exhibit the Painlev\'{e} property for any value of the parameter $%
\lambda $. We present analytic solutions expressed in terms of Right Laurent
expansions for various values of the parameter $\lambda $. Finally, we
discuss the qualitative evolution of the effective equation of state
parameter for these analytic solutions.
\end{abstract}

\maketitle

\section{Introduction}

\label{sec1}

In gravitational physics, exact and analytic solutions are of particular
interest because they enable precise modeling of gravitational fields and
spacetime structures \cite{steph1}. Such solutions describe strong
gravitational fields in black hole spacetimes \cite{sl1}, the early-time
acceleration phase of the universe \cite{sl4,sl5,sl6}, dark energy models 
\cite{sl7,sl8,sl9}, cosmological singularities, and many other phenomena 
\cite{sl10,sl11,sl12,sl13,sl14,sl15,sl16}. Furthermore, exact and analytic
solutions serve as reference points for evaluating numerical simulations and
approximation techniques, particularly near singularities. For instance, the
exact Kasner solution approximates the behaviour of the Mixmaster universe
near the cosmological singularity \cite{sl17}. Additionally, these solutions
provide insights into the initial value problem and are essential for
reducing the computational cost of numerical simulations. For further
details, we refer the reader to the discussion in \cite{steph2}.

In cosmological studies, scalar fields are of particular interest as they
introduce new degrees of freedom that help describe various physical
phenomena. The early acceleration phase of the universe is attributed to the
presence of an inflationary field \cite{sl4}. Moreover, scalar fields can
account for late-time cosmic acceleration and explain the nature of dark
energy \cite{Ratra,Barrow,q14,q15,cop,orl1,orl2,orl3,orl4}. Additionally,
they can encapsulate the geometric degrees of freedom introduced by modified
theories of gravity \cite{sot}.

The cosmological field equations for minimally coupled scalar field models
depend on a potential function that defines the mass component of the scalar
field and drives its dynamics. When this potential term dominates, the
scalar field mimics the cosmological constant \cite{Barrow}. However, in
general, the scalar field potential does not necessarily dominate, meaning
that the future attractor may differ from that of a de Sitter universe \cite%
{cop}. The choice of scalar potential determines the behaviour of cosmic
parameters throughout cosmic evolution. Asymptotic methods \cite%
{q15,cop,gn1,gn2,gn3} are often employed to study the main phases of cosmic
evolution dictated by a given model. The determination of analytic solutions
is crucial, as it provides valuable insights into the model's behaviour and
its initial value problem. It is noteworthy that integrable dynamical
systems are insensitive to small changes in initial conditions, implying the
absence of chaotic trajectories in the solution space.

There are relatively few analytic solutions in the literature for
nonminimally coupled scalar fields. The complete analytic solution for the
exponential potential $V\left( \phi \right) =V_{0}e^{\lambda \phi }$ was
presented in \cite{Russo}. By applying techniques from analytical mechanics,
new analytic solutions for hyperbolic and exponential potentials have been
derived \cite{Basilakos,an1,an2,an3,an4,an5,an6}. A comprehensive list of
integrable scalar field potentials can be found in \cite{l1}, while
integrable cosmological models incorporating additional matter sources are
presented in \cite{l2}.

Recently, the Eisenhart-Duval lift \cite{ll01,ll02,ll03} was applied to the
analysis of the cosmological field equations in scalar field cosmology \cite%
{lift1}. It was found that, for the exponential potential, the cosmological
field equations can be linearized via canonical transformations in the
extended space, allowing the analytic solution to be expressed in terms of a
free particle. A similar approach has been employed to study other
gravitational models, such as field equations in static spherically
symmetric spacetimes and anisotropic cosmologies \cite{lift2}. However, the
canonical transformation of the extended space into the original space
results in a nonlocal transformation. Consequently, local conservation laws
in the extended space can be used to construct nonlocal conservation laws
for the original system. In this study, we apply this property to determine
scalar field potentials that satisfy nonlocal conservation laws. For further
applications of the Eisenhart-Duval lift in gravitational physics, we refer
the reader to \cite{ll04,ll05,ll06}. The structure of the paper is as
follows.

In Section \ref{sec2}, we present the gravitational model under
consideration. Specifically, we examine a spatially flat Friedmann--Lema%
\^{\i}tre--Robertson--Walker geometry, where the cosmological fluid consists
of a dust fluid and a scalar field minimally coupled to gravity. The
gravitational field equations form a singular Hamiltonian system that
depends on the scalar field potential and the nature of the scalar field. We
assume the scalar field to be either quintessence or a phantom scalar field.

In Section \ref{sec3}, we discuss an approach for determining nonlocal
conservation laws for the dynamical system of our gravitational model. We
employ the Eisenhart-Duval lift and express the field equations in an
equivalent form as a set of geodesic equations for an extended
minisuperspace. Consequently, the problem of determining conservation laws
and symmetries for the original system reduces to the classification of the
Conformal Killing symmetries of the extended minisuperspace. We introduce a
novel lift and solve the classification problem for the Conformal Killing
symmetries. We find that for the scalar field potential $V\left( \phi
\right) =\alpha \left( e^{\lambda \phi }+\beta \right)$, the geodesic
equations possess additional conservation laws, which manifest as nonlocal
conservation laws for the cosmological model.

The admitted nonlocal conservation laws do not satisfy all the necessary
conditions to determine the integrability of the dynamical system, in
Section \ref{sec4}, we apply the Painlev\'{e} analysis to study the
integrability of this cosmological model. We outline the basic steps of the
ARS algorithm and apply it to both the quintessence and phantom scalar field
cases. Finally, in Section \ref{sec5}, we summarize our conclusions.

\section{Scalar field FLRW Cosmology}

\label{sec2}

We begin our model by introducing the spatially flat FLRW geometry as a
framework to describe the universe, expressed through the metric tensor $%
g_{\mu \nu }$ with line element%
\begin{equation}
ds^{2}=-N^{2}\left( t\right) dt^{2}+a^{2}\left( t\right) \left(
dx^{2}+dy^{2}+dz^{2}\right) ,  \label{sf.01}
\end{equation}%
where $N\left( t\right) $ is the lapse function and $a\left( t\right) $ is
the scale factor, represents the radius of the universe. We define the
comoving observer $u^{\mu }=\frac{1}{N}\delta _{t}^{\mu }$, that is, $u^{\mu
}u_{\mu }=-1$. Thus, the kinetic quantities of shear~$\sigma _{\mu \nu }$,
vorticity~$\omega _{\mu \nu }$ and expansion~$\theta $ for the line element (%
\ref{sf.01}) are given%
\begin{equation}
\sigma =0,~\omega _{\mu \nu }=0\text{ and }\theta =3H,  \label{ss.02}
\end{equation}%
where $H$ is the Hubble function defined as 
\begin{equation}
H=\frac{1}{N}\frac{d\ln a}{dt}.  \label{ss.03}
\end{equation}

Within the framework of General Relativity, we consider that the
cosmological fluid is consisted by the dark matter and the dark energy
without any energy transfer between them. For the dark matter it is
described by dust fluid source with energy momentum 
\begin{equation}
T_{\mu \nu }^{DM}=\rho _{DM}u_{\mu }u_{\nu }.  \label{ss.04}
\end{equation}%
For the dark energy we consider the energy momentum tensor%
\begin{equation}
T_{\mu \nu }^{DE}=\left( \rho _{DE}+p_{DE}\right) u_{\mu }u_{\nu
}+p_{DE}g_{\mu \nu },  \label{ss.05}
\end{equation}%
which follows from the variation of the Lagrangian function for the
quintessence field\qquad\ 
\begin{equation}
\mathcal{L}\left( \phi ,\phi _{;\kappa }\right) =-\frac{1}{2}g^{\mu \nu
}\phi _{;\mu }\phi _{;\nu }+V(\phi ),  \label{sf.02}
\end{equation}%
with respect to the metric tensor $T_{\mu \nu }^{DE}=\frac{\partial \left( 
\sqrt{-g}L^{\phi }\right) }{\partial g_{\mu \nu }}$, that is,%
\begin{eqnarray}
\rho _{DE} &=&\frac{\varepsilon }{2}\frac{\dot{\phi}^{2}}{N^{2}}+V(\phi ),
\label{ss.06} \\
p_{DE} &=&\frac{\varepsilon }{2}\frac{\dot{\phi}^{2}}{N^{2}}-V(\phi ).
\label{ss.07}
\end{eqnarray}%
where $\varepsilon =\pm 1$. Parameter $\varepsilon =1$ corresponds to the
quintessence scalar field \cite{Ratra}, while $\varepsilon =-1$ to the
phantom field~\cite{q14}.

The equation of state parameter for the dark energy fluid is defined as 
\begin{equation}
w_{DE}=\frac{p_{DE}}{\rho _{DE}}=\frac{\varepsilon \dot{\phi}%
^{2}-2N^{2}V(\phi )}{\varepsilon \dot{\phi}^{2}+2N^{2}V(\phi )}.
\label{ss.08}
\end{equation}%
Therefore, for the quintessence scalar field the equation of state parameter
is bounded $-1\leq w_{DE}\leq 1$. The upper limit occurs when the kinetic
term of the scalar field prevails $\dot{\phi}^{2}>>V\left( \phi \right) $,
while the lower limit is reached when the potential term dominates, i.e. $%
\dot{\phi}^{2}<<V\left( \phi \right) $. On the other hand, for the phantom
scalar field there is not a lower bound for the equation of state parameter 
\cite{q15}.

The evolution of the physical parameters is given by the Einstein's field
equations, they are%
\begin{eqnarray}
3H^{2} &=&\rho _{DM}+\rho _{DE},  \label{ss.10} \\
\frac{2}{N}\dot{H}+3H &=&-p_{DE}.  \label{ss.11}
\end{eqnarray}%
Moreover, the equation of motion for the scalar field is 
\begin{equation}
\frac{\varepsilon }{N}\left( \frac{d}{dt}\left( \frac{1}{N}\dot{\phi}\right)
+3H\dot{\phi}\right) +V_{,\phi }=0,  \label{ss.12}
\end{equation}%
which the equation of motion of the dust fluid source is%
\begin{equation}
\frac{1}{N}\dot{\rho}_{DM}+3H\rho _{DM}=0.  \label{ss.14}
\end{equation}%
It follows from the latter equation that, $\rho _{DM}\left( a\right) =\rho
_{m0}a^{-3}$, where $\rho _{m0}$ is an integration constant.

The dynamical quantities of the cosmological problem are the lapse function $%
N\left( t\right) $, the scale factor, $a\left( t\right) $ and the scalar
field $\phi \left( t\right) $. The second-order differential equations (\ref%
{ss.11}), (\ref{ss.12}) describe the evolution of the scale factor and of
the scalar field, while the first-order differential equation (\ref{ss.10})
provides the algebraic constraint for the lapse function $N\left( t\right) $%
. The definition of the scalar field potential $V\left( \phi \right) $
provides the dynamics for the cosmological fluid and provides the physical
properties of the model.

The cosmological field equations (\ref{ss.10}), (\ref{ss.11}), (\ref{ss.12}%
)\ form a dynamical system which follows from the variation of the singular
point-like Lagrangian%
\begin{equation}
L\left( N,a,\dot{a},\phi ,\dot{\phi}\right) =\frac{1}{N}\left( -3a\dot{a}%
^{2}+\frac{\varepsilon }{2}a^{3}\dot{\phi}^{2}\right) -N\left( a^{3}V\left(
\phi \right) +\rho _{m0}\right) \text{.}  \label{ss.15}
\end{equation}%
The constraint (\ref{ss.10}) follows from variation with respect to the
lapse function $N\left( t\right) $, while the second-order differential
equations follows from the variation with respect to the dynamical variables 
$a\left( t\right) $ and $\phi \left( t\right) $.

In the following, without loss of generality we assume the lapse function to
be constant, that is, $N\left( t\right) =1$. We introduce the momentum~$%
p_{a}=-6a\dot{a},~p_{\phi }=a^{3}\dot{\phi}$, and we write the Hamiltonian
function%
\begin{equation}
\mathcal{H}\equiv \frac{1}{2}\left( -\frac{p_{a}^{2}}{6a}+\frac{p_{\phi }^{2}%
}{\varepsilon a^{3}}\right) +\left( a^{3}V\left( \phi \right) +\rho
_{m0}\right) =0,  \label{ss.16}
\end{equation}%
which describes the cosmological model. The field equations in terms of the
Hamiltonian formalism are%
\begin{eqnarray}
\dot{a} &=&-\frac{1}{6}\frac{p_{a}}{a}, \\
\dot{\phi} &=&\frac{p_{\phi }}{\varepsilon a^{3}}, \\
\dot{p}_{a} &=&-\frac{1}{2}\left( \frac{p_{a}^{2}}{6a^{2}}-3\frac{p_{\phi
}^{2}}{a^{4}}\right) -3a^{2}V\left( \phi \right) , \\
\dot{p}_{\phi } &=&-a^{3}V_{,\phi }.
\end{eqnarray}

\section{The Eisenhart-Duval lift}

\label{sec3}

We introduce the extended Hamiltonian function \cite{ani0}%
\begin{equation}
\mathcal{H}_{lift}\equiv \frac{1}{2}\left( -\frac{p_{a}^{2}}{6a}+\frac{%
p_{\phi }^{2}}{\varepsilon a^{3}}\right) +\frac{1}{2}p_{z}^{2}+a^{3}V_{1}%
\left( \phi \right) p_{u}^{2}+a^{3}V_{2}\left( \phi \right) p_{u}p_{v}=0,
\label{ss.17}
\end{equation}%
which describes the null geodesic equations for the Riemannian manifold with
line element%
\begin{equation}
d\hat{s}_{lift}^{2}=-6ada^{2}+\varepsilon a^{3}d\phi ^{2}+dz^{2}+\frac{2}{%
a^{3}V_{2}\left( \phi \right) }dudv-2\frac{V_{1}\left( \phi \right) }{%
a^{3}\left( V_{2}\left( \phi \right) \right) ^{2}}du^{2}.  \label{ss.18}
\end{equation}%
The momentum $p_{a},~p_{\phi },~p_{u},$ $p_{v}$ and $p_{z}$ are defined 
\begin{eqnarray}
\dot{a} &=&-\frac{1}{6}\frac{p_{a}}{a}, \\
\dot{\phi} &=&\frac{p_{\phi }}{a^{3}}, \\
\dot{u} &=&a^{3}V_{2}\left( \phi \right) p_{v}, \\
\dot{v} &=&a^{3}\left( V_{2}\left( \phi \right) p_{u}+2V_{1}\left( \phi
\right) p_{v}\right) , \\
\dot{z} &=&p_{z}.~
\end{eqnarray}%
Moreover, the rest of the Hamilton's equations are%
\begin{eqnarray}
\dot{p}_{a} &=&-\frac{1}{2}\left( \frac{p_{a}^{2}}{6a^{2}}-3\frac{p_{\phi
}^{2}}{\varepsilon a^{4}}\right) -3a^{2}\left( V_{1}\left( \phi \right)
p_{u}^{2}+V_{2}\left( \phi \right) p_{u}p_{v}\right) , \\
\dot{p}_{\phi } &=&-a^{3}\left( V_{1,\phi }p_{u}^{2}+V_{2,\phi
}p_{u}p_{v}\right) , \\
\dot{p}_{u} &=&0, \\
\dot{p}_{v} &=&0, \\
\dot{p}_{z} &=&0\text{.}
\end{eqnarray}%
For arbitrary functions $V_{1}\left( \phi \right) ,~V_{1}\left( \phi \right) 
$ \ the geodesic equations admits the conservation laws $p_{z}=p_{z}^{0},$ $%
p_{u}=p_{u}^{0}$ and $p_{v}=p_{v}^{0}$. These conservation laws are
constructed by the three translations $\partial _{z},~\partial _{v}$ and $%
\partial _{v}$, which are Killing symmetries for the line element (\ref%
{ss.18}).

The application of these conservation laws in (\ref{ss.17}) gives the
Hamiltonian function (\ref{ss.16}) with the constraints $p_{z}=\sqrt{2\rho
_{m0}}$ and 
\begin{equation}
V\left( \phi \right) =V_{1}\left( \phi \right) \left( p_{u}^{0}\right)
^{2}+V_{2}\left( \phi \right) p_{u}^{0}p_{v}^{0}.  \label{ss.19}
\end{equation}%
Without loss of generality we can absorb the integration constants inside
the potential function, thus we consider $p_{u}^{0}=1$ and $p_{v}^{0}=1$.

However, for specific forms of function $V_{1}\left( \phi \right) $ and $%
V_{2}\left( \phi \right) $, the null geodesic with Hamiltonian (\ref{ss.17})
admits additional conservation laws related to the Conformal Killing
symmetries of (\ref{ss.18}).

We assume that $V_{,\phi }\neq 0$, therefore for $V_{1}\left( \phi \right)
=\alpha e^{\lambda \phi }$ and $V_{2}\left( \phi \right) =\alpha \beta $,
that is, for the scalar field potential function 
\begin{equation}
V\left( \phi \right) =\alpha \left( e^{\lambda \phi }+\beta \right) ,
\label{ss.20}
\end{equation}%
the geodesic equations admits the conservation laws%
\begin{eqnarray}
I_{1} &=&\frac{1}{3}ap_{a}+up_{u}+vp_{v}+\frac{1}{2}zp_{z},  \label{ss.21} \\
I_{2} &=&\frac{\lambda }{6}ap_{a}+\varepsilon p_{\phi }+\lambda vp_{v}+\frac{%
\lambda }{4}zp_{z}.  \label{ss.22}
\end{eqnarray}%
The corresponding symmetry vectors are~$\frac{a}{3}\partial _{a}+u\partial
_{u}+v\partial _{v}+\frac{z}{2}\partial _{z}$ and $\partial _{\phi }+\frac{%
\lambda }{2}\left( \frac{a}{3}\partial _{a}+v\partial _{v}+\frac{z}{2}%
\partial _{z}\right) $. \ 

The case with $\beta =0$, corresponds to the pure exponential potential
which it is known that it is integrable. It is analytic solution have been
found recently with the use of the Einsenhart-Duval lift in \cite{lift1}.
Hence, in the following we assume $\alpha \beta \neq 0$

\subsection{Nonlocal symmetries in Scalar field Cosmology}

Although the conservation laws $I_{1}$ and $I_{2}$ are local for the null
geodesic equations, they are nonlocal for the original dynamical system (\ref%
{ss.16}). Since $p_{u}=1,$ $p_{v}=1$ and $p_{z}=\sqrt{2\rho _{m0}}$ the
conservation laws are simplified as%
\begin{eqnarray}
I_{1} &=&\frac{1}{3}ap_{a}+\sqrt{\frac{\rho _{m0}}{2}}t+2\alpha \int \left(
e^{\lambda \phi }+\beta \right) a^{3}dt, \\
I_{2} &=&\frac{\lambda }{2}\left( \frac{a}{3}p_{a}+\sqrt{\frac{\rho _{m0}}{2}%
}t\right) +\varepsilon p_{\phi }+\lambda \int \left( \alpha \left(
2e^{\lambda \phi }+\beta \right) \right) a^{3}dt.
\end{eqnarray}%
Moreover, the conserved quantity $\hat{I}=I_{1}-\frac{2}{\lambda }I_{2}$ is
time independent, that is,%
\begin{equation}
\hat{I}\equiv \frac{2}{\lambda }p_{\phi }-2\alpha \int e^{\lambda \phi
}a^{3}dt.
\end{equation}

By using the conservation laws $\left\{
H,I_{1},I_{2},p_{u},p_{v},p_{z}\right\} $, the ten equations which describe
the null geodesic equations are reduced to the following five first order
differential equations%
\begin{eqnarray}
\dot{x} &=&\frac{3}{2}\left( u+v-I_{1}+\sqrt{\frac{\rho _{m0}}{2}}z\right) ,
\\
\dot{\phi} &=&\frac{2I_{2}-I_{1}\lambda +\lambda \left( u-v\right) }{2x}, \\
\dot{u} &=&\alpha \beta x, \\
\dot{v} &=&\varepsilon \left( \frac{3}{2}\frac{\dot{x}^{2}}{x}-2\rho
_{m0}-\alpha \beta x\right) -\frac{\left( I_{1}\lambda -2I_{2}-\lambda
\left( u-v\right) \right) ^{2}}{4x}, \\
\dot{z} &=&\sqrt{2\rho _{m0}},
\end{eqnarray}%
where $x^{3}=a$. This five dimensional system is not separable, that is not
a real surprise, because not all the conservation laws $\left\{
H,I_{1},I_{2},p_{u},p_{v},p_{z}\right\} $ are in involution. Consequently,
we can not infer about the Liouville integrability of the dynamical system
by using these conservation laws.

The five dimensional system can be reduced to the following
three-dimensional system%
\begin{eqnarray}
\frac{1}{3}\ddot{x}+\lambda \left( x\Phi \right) ^{\cdot }-\alpha \beta -%
\frac{1}{2}\rho _{m0} &=&0,  \label{ss.24} \\
\frac{1}{3}\left( \ddot{x}+2\frac{\dot{x}^{2}}{x}\right) -\lambda \left(
x\Phi \right) ^{\cdot }+x\left( \alpha \beta +\varepsilon \Phi ^{2}\right) +%
\frac{3}{2}\rho _{m0} &=&0.  \label{ss.25}
\end{eqnarray}%
where now $\dot{\phi}=\Phi $. Dynamical system (\ref{ss.24}), (\ref{ss.25})
describes the second-order field equations (\ref{ss.11}), (\ref{ss.12})
after the application of the constraint equation.

We continue our analysis with the investigation of the stability properties
for this dynamical system by using the Painlev\'{e} analysis.

\section{Painlev\'{e} analysis}

\label{sec4}

The Painlev\'{e} analysis, also known as singularity analysis, of
differential equations is a fundamental tool for determining the
integrability of dynamical systems. The method has been widely applied for
the study of the integrability properties of various cosmological models,
see for instance \cite{sin1,sin2,sin3,sin4,sin5,sin6,sin7,sin8} and
references.

Nowadays, the modern treatments of the singularity analysis is typically
summarized by the ARS algorithm, proposed by Ablowitz, Ramani, and Segur 
\cite{Abl1,Abl2,Abl3}. Within the following lines, we present the three
basic steps of the ARS algorithm for ordinary differential equations. For a
detailed review we refer the reader to \cite{buntis}.

Assume that $\mathcal{F}\left( t,x\left( t\right) ,\dot{x}\left( t\right) ,%
\ddot{x}\left( t\right) ,...\right) \equiv 0$ describes a dynamical system,
where $t$ is the independent variable and $x\left( t\right) $ is the
dependent variable. In the first step of the ARS algorithm, we determine
whether a movable singularity exists. To achieve this, we investigate
whether, near the singularity, the solution of the differential equation is
asymptotically described by the function 
\begin{equation}
x\left( t\right) =x_{0}\left( t-t_{0}\right) ^{p},  \label{nn.01}
\end{equation}%
where $t_{0}$ denotes the position of the singularity, which serves as an
integration constant in autonomous dynamical systems. By substituting (\ref%
{nn.01}) into the dynamical system and considering the dominant power terms,
we derive the values of the two variables, $p$ and $x_{0}$. The exponent $p$
should be a negative number for the singularity to be a movable pole.
However, this condition has been relaxed to include fractional and even
positive exponents, since the derivative of a positive fractional exponent
eventually results in a negative exponent, leading to a singularity.

The second step involves determining the resonances, which correspond to the
positions of the integration constants within the expansion. We substitute $%
x=a(t-t_{0})^{p}+m(t-t_{0})^{p+S}$ into the dominant terms of the equation
and collect the terms linear in $m$, as this is where the coefficient first
appears in the expansion. If the multiplier of $m$ is zero, the value of $m$
is arbitrary. The coefficient is a polynomial in the resonances $S$, whose
solutions provide the specific resonance values. The number of resonances
should equal the order of the dynamical system. One of these resonances must
be $-1$, which corresponds to the movable singularity at $t_{0}$.

Finally, if the dynamical system satisfies the previous steps, we substitute
an appropriate Laurent series into the full ordinary differential equation
to verify consistency with the type of series suggested by the dominant term
analysis.

If all these conditions are met, the equation under study is said to possess
the Painlev\'{e} property (weak in the case of fractional exponents) and is
conjectured to be integrable. In this case, at least one viable Laurent
expansion exists as a solution.

The ARS algorithm is applied to investigate if the dynamical system (\ref%
{ss.24}), (\ref{ss.25}) possesses the Painlev\'{e} property and if the
analytic solution can be expressed in terms of a Laurent expansion. We
perform the analysis for the quintessence field, $\varepsilon =+1$ and the
phantom field, $\varepsilon =-1$.

\subsection{Quintessence scalar field}

For the quintessence scalar field, i.e. $\varepsilon =+1,$with potential (%
\ref{ss.20}), the field equations are described by the dynamical system (\ref%
{ss.24}), (\ref{ss.25}) the ARS algorithm gives the leading order behaviour 
\begin{eqnarray}
x_{q}\left( t\right) &=&x_{0}\left( t-t_{0}\right) ^{\frac{6}{\lambda ^{2}}%
},~ \\
\Phi _{q}\left( t\right) &=&-\frac{2}{\lambda }\left( t-t_{0}\right) ^{-1}
\end{eqnarray}%
with resonances%
\begin{equation}
S=-1,~S=0\text{, }S=1-\frac{6}{\lambda ^{2}}\text{.}
\end{equation}%
where $x_{0},~t_{0}$ are two integration constants. The third integration
constant is located in the position of the third resonance, that is, it
depends on parameter $\lambda $. The leading order behaviour is that of the
scaling solution of the exponential scalar field. For $\lambda ^{2}<6$, the
third resonance is negative which means that the solution is described by a
Left Laurent expansion, while for $\lambda ^{2}>6$, the analytic solution is
expressed by a Right Laurent expansion.

We consider $\lambda =2\sqrt{3}$, such that the third resonance read $S=%
\frac{1}{2}$. Thus, the analytic solution of the model is given by the
following by the following expansions 
\begin{eqnarray}
x\left( t\right) &=&x_{0}\left( t-t_{0}\right) ^{\frac{1}{2}}+x_{1}\left(
t-t_{0}\right) +x_{2}\left( t-t_{0}\right) ^{\frac{1+2}{2}}+...~, \\
\Phi \left( t\right) &=&-\frac{2}{\lambda }\left( t-t_{0}\right) ^{-1}+\phi
_{1}\left( t-t_{0}\right) ^{-\frac{1}{2}}+\phi _{2}\left( t-t_{0}\right) ^{%
\frac{2-2}{2}}+...~,
\end{eqnarray}%
where the third integration constant is variable $x_{1}$, and 
\begin{equation}
x_{2}=\frac{7}{64}\frac{x_{1}^{2}}{x_{0}},~x_{3}=-\frac{7x_{1}^{3}}{%
160x_{0}^{2}}+\frac{2}{5}\rho _{m0},~x_{4}=\frac{651}{40960}\frac{x_{1}^{4}}{%
x_{0}^{3}}+\frac{7}{40}\frac{x_{1}}{x_{0}}\rho _{m0}+\frac{2}{3}\alpha \beta
x_{0},~...
\end{equation}%
and%
\begin{equation}
\Phi _{2}=\frac{x_{1}^{2}}{\sqrt{3}32x_{0}^{2}},~\Phi _{3}=\frac{163}{\sqrt{3%
}1250}\frac{x_{1}^{3}}{x_{0}^{2}}+\frac{3\sqrt{3}}{5x_{0}}\rho _{m0},~\Phi
_{4}=-\frac{1051x_{1}^{4}}{5120\sqrt{3}x_{0}^{4}}-\frac{4\sqrt{3}}{5}\frac{%
x_{1}}{x_{0}^{2}}\rho _{m0}+\frac{4}{3\sqrt{3}}\alpha \beta ,~...
\end{equation}

In terms of physically accepted solutions, the leading-order behaviour
describes a scaling solutions with equation of state parameter $w_{l}\left(
t\right) =-1+\frac{\lambda ^{2}}{3}$, thus the condition $\left\vert
w_{l}\left( t\right) \right\vert \leq $ $1$ constraint parameter $\lambda $
as $\lambda ^{2}<6$. Thus, this singular solution describes physical
solution far from the singularity, that is, for $t-t_{0}>>1$.

\begin{figure}[tbp]
\includegraphics[width=0.5\textwidth]{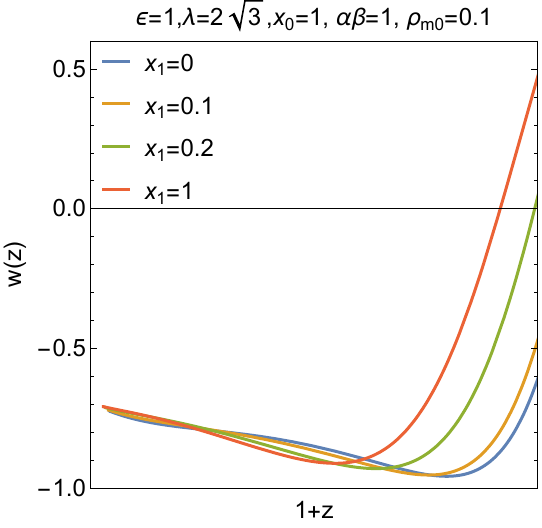}
\caption{Evolution of the equation of state parameter $w\left( a\right) $
for the analytic solution given by the Right Laurent expansion for~the
quintessence scalar field and $\protect\lambda =2\protect\sqrt{3}$ for
different values of the free parameters.}
\label{fig1}
\end{figure}

In Fig. \ref{fig1} we present the qualitative evolution of the equation of
state parameter for the first eleventh term of the Laurent expansion for the
quintessence scalar field and~$\lambda =2\sqrt{3}$. It is important to
mention that the accuracy of the solution is improved as we introduce
additional terms in the expansion.

On the other hand, for $\lambda ^{2}<6\,\ $the third step of the ARS
algorithm fails. We conclude that the dynamical system does not possess the
Painlev\'{e} property in this case.

It is important to mention that for the case $\lambda ^{2}>6,$ the leading
dominant term has positive exponent and the resonances are positive. Thus,
as we move far from the singularity the rest of the terms dominates.

\subsection{Phantom scalar field}

Consider now the phantom scalar field, i.e. $\varepsilon =-1$, then for the
potential (\ref{ss.20}), the field equations are described by the dynamical
system (\ref{ss.24}), (\ref{ss.25}). The application of the ARS algorithm
gives the leading order terms%
\begin{eqnarray}
x_{p}\left( t\right) &=&x_{0}\left( t-t_{0}\right) ^{-\frac{6}{\lambda ^{2}}%
},~ \\
\Phi _{p}\left( t\right) &=&\frac{2}{\lambda }\left( t-t_{0}\right) ^{-1}
\end{eqnarray}%
with resonances%
\begin{equation}
S=-1,~S=0\text{, }S=1+\frac{6}{\lambda ^{2}}.
\end{equation}%
Thus, the location of the singularity, $t_{0}$, coefficient $x_{0}$ are the
two integration constants, while the third integration constant it depend on
parameter $\lambda $. Speciality the location in the Laurent expansion is
given by the third resonance. Because the three resonances are always
positive the solutions are expressed by Right Laurent expansions.

We consider $\lambda =1$, and we write the solution%
\begin{eqnarray*}
x\left( t\right) &=&x_{0}\left( t-t_{0}\right) ^{-6}+x_{1}\left(
t-t_{0}\right) ^{-5}+x_{2}\left( t-t_{0}\right) ^{-4}+...~, \\
\Phi \left( t\right) &=&2\left( t-t_{0}\right) ^{-1}+\Phi _{1}+\Phi
_{2}\left( t-t_{0}\right) +...~,
\end{eqnarray*}%
where now the third integration constant is $x_{7}$ and%
\begin{equation*}
x_{1}=0,~x_{2}=-\frac{\alpha \beta x_{0}}{5},~x_{3}=0,~x_{4}=\frac{17x_{0}}{%
600}\left( \alpha \beta \right) ^{2},~x_{5}=0,
\end{equation*}%
\begin{equation*}
~x_{6}=-\frac{1049}{189000}x_{0}\left( \alpha \beta \right) ^{3},~x_{8}=%
\frac{\rho _{m0}}{48}-\frac{1357}{1512~}\left( \frac{\alpha \beta }{10}%
\right) ^{4}x_{0},~...
\end{equation*}%
and%
\begin{equation*}
\Phi _{1}=0,~\Phi _{2}=-\frac{\alpha \beta }{15},~\Phi _{3}=0,~\Phi _{4}=%
\frac{7}{450}\left( \alpha \beta \right) ^{2},~\Phi _{5}=0,~\Phi _{6}=-\frac{%
289}{2365}\left( \alpha \beta \right) ^{3},~
\end{equation*}%
\begin{equation*}
\Phi _{7}=\frac{7x_{7}}{x_{0}},~\Phi _{8}=-\frac{2699\left( \alpha \beta
\right) ^{4}x_{0}-140000\rho _{m0}}{315000x_{0}},~\Phi _{9}=\frac{21\alpha
\beta }{10x_{0}}x_{7},~...
\end{equation*}

\begin{figure}[tbp]
\includegraphics[width=1\textwidth]{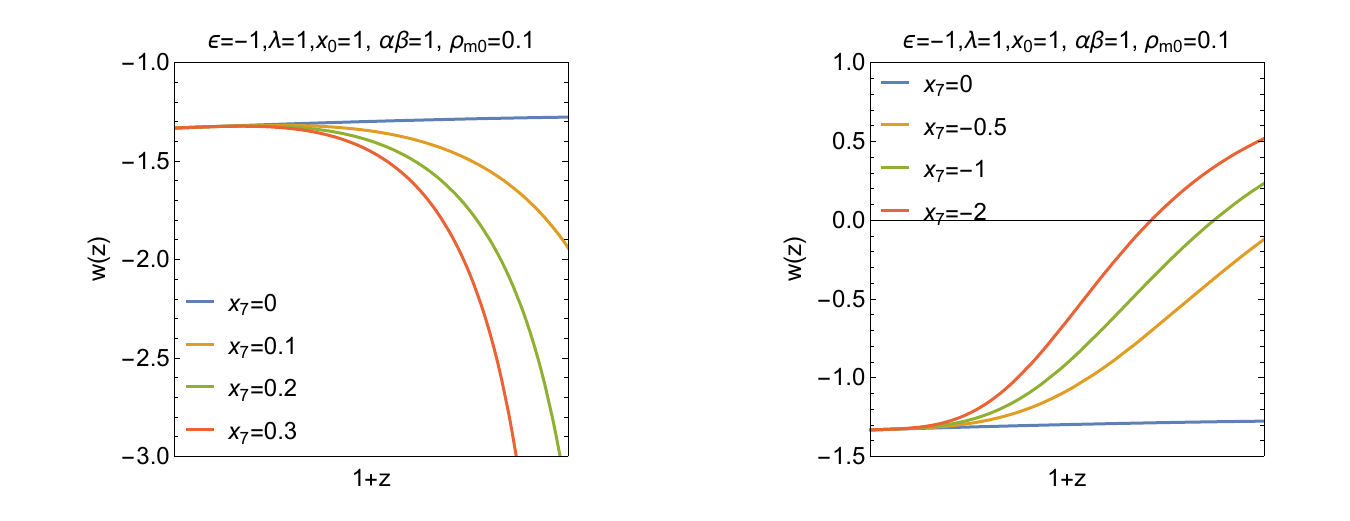}
\caption{Evolution of the equation of state parameter $w\left( a\right) $
for the analytic solution given by the Right Laurent expansion for~the
phantom scalar field and $\protect\lambda =1$ for different values of the
free parameters.}
\label{fig2}
\end{figure}

In Fig. \ref{fig2} we present the qualitative evolution of the equation of
state parameter for the first eleventh term of the Laurent expansion for the
phantom scalar field for $\lambda =1$. It is important to mention that the
accuracy of the solution is improved as we introduce additional terms in the
expansion.

From Fig. \ref{fig2} we observe two different behaviours for the effective
equation of state which depend on the value of parameter $x_{7}$. For $%
x_{7}>0$, the $w\left( a\right) $ originates from negative infinity and
approaches a scaling solution, while for $x_{7}<0$ the equation of state
parameter starts from positive values and converges to the attractor. Recall
that in both cases, the attractor is described by the leading term $a\left(
t\right) \simeq \left( t-t_{0}\right) ^{-2}$.

Moreover, we test the solution provided by the ARS algorithm and for the
case where $\lambda =2\sqrt{3},$we find 
\begin{eqnarray*}
x\left( t\right) &=&x_{0}\left( t-t_{0}\right) ^{-\frac{1}{2}%
}+x_{1}+x_{2}\left( t-t_{0}\right) ^{\frac{1}{2}}+~...~, \\
\Phi \left( t\right) &=&\frac{1}{\sqrt{3}}\left( t-t_{0}\right) ^{-1}+\Phi
_{1}\left( t-t_{0}\right) ^{-\frac{1}{2}}+\Phi _{2}+~...~,
\end{eqnarray*}%
then third integration constant is the coefficient $x_{3}$, while for the
rest of the coefficients it follows%
\begin{equation*}
x_{1}=0,~x_{2}=0,~x_{4}=2\alpha \beta x_{0},~x_{5}=\frac{6}{35}\rho
_{m0},~x_{6}=\frac{27}{128}\frac{x_{3}^{2}}{x_{0}},~...~,
\end{equation*}%
and%
\begin{equation*}
\Phi _{1}=0,~\Phi _{2}=0,~\Phi _{3}=\frac{\sqrt{3}}{4}\frac{x_{3}}{x_{0}}%
,~\Phi _{4}=\frac{4\sqrt{3}}{3}\alpha \beta ,~\Phi _{5}=\frac{5\sqrt{3}}{7}%
\frac{\rho _{m0}}{x_{0}},~\Phi _{6}=\frac{43\sqrt{3}}{64}\left( \frac{x_{3}}{%
x_{0}}\right) ^{2},~...~.
\end{equation*}

We conclude that for the phantom scalar field the field equations pass the
three tests of the ARS algorithm for arbitrary value of parameter $\lambda $.

\section{Conclusions}

\label{sec5}

In this study, we employed the Eisenhart-Duval lift and the Painlev\'{e}
analysis to address the problem of integrability for the field equations in
scalar field cosmology. Within the spatially flat FLRW background geometry,
we introduced a minimally coupled scalar field, either quintessence or
phantom, along with a dust fluid source. We made use of the minisuperspace
description to express the system as an equivalent set of geodesic equations
in an extended minisuperspace.

The structure of the extended minisuperspace depends on the scalar field
potential. Thus, the requirement that the geodesic equations admit
conservation laws is equivalent to the existence of a nontrivial conformal
algebra in the extended space. Consequently, we determined the functional
forms of the scalar field potential for which conformal symmetries exist in
the extended minisuperspace. This condition led to the identification of two
scalar field potentials: the pure exponential potential, a well-known
integrable model, and the exponential potential with an additional
cosmological constant term.

For the latter potential, we derived the local conservation laws in the
extended minisuperspace, which correspond to nonlocal conservation laws in
the original dynamical system. However, the existence of conservation laws
does not necessarily ensure the integrability of the field equations. To
further investigate this, we applied the Painlev\'{e} analysis. Our results
show that the phantom scalar field always leads to an integrable set of
field equations, whereas for the quintessence scalar field, the field
equations exhibit the Painlev\'{e} property only when the two resonances
obtained via the ARS algorithm are positive.

This new result extends the analysis of integrable models in cosmological
theories. Furthermore, the methodology developed in this work opens new
directions for the application of the Eisenhart-Duval lift and the Painlev%
\'{e} analysis in modern cosmology.

\begin{acknowledgments}
AP thanks the support of VRIDT through Resoluci\'{o}n VRIDT No. 096/2022 and
Resoluci\'{o}n VRIDT No. 098/2022.
\end{acknowledgments}


\begin{thebibliography}{99}
\bibitem{steph1} H. Stephani, D. Kramer M. MacCallum, C. Hoenselaers and E.
Herlt, Exact Solutions to Einstein's Field Equations, Second Edition,
Cambrigde University Presss, Cambridge (2003)

\bibitem{sl1} V.P. Frolov and A.\ Zelnikov, Introduction to Black Hole
Physics, Oxford University Press, Oxford (2012)

\bibitem{sl4} A.D. Linde, Phys. Lett. B 129, 177 (1983)

\bibitem{sl5} J. Martin, C. Ringeval, R. Trotta and V. Vennin, JCAP 03, 039
(2014)

\bibitem{sl6} S.V. Chernov, Phys. Lett. B 398, 269 (1997)

\bibitem{sl7} C. Gao and Y.-G. Shen, Gen.\ Rel. Grav. 48, 131 (2016)

\bibitem{sl8} P. Christodoulidis and A. Paliathanasis, JCAP 05, 038 (2021)

\bibitem{sl9} A. Paliathanasis and M. Tsamparlis, Phys.\ Rev. D 60, 043529
(2014)

\bibitem{sl10} P. Szekeres, Commun. Math. Phys. 41, 55 (1975)

\bibitem{sl11} A. Krasinski, Inhomogeneous Cosmological Models, Cambridge
University Press, Cambridge, (1997)

\bibitem{sl12} D.A. Szafron, J. Math. Phys. 18, 1673 (1977)

\bibitem{sl13} S.D. Odintsov and V.K. Oikonomous, Phys.\ Rev. D 107, 104039
(2023)

\bibitem{sl14} S. Nojiri and S.D. Odintsov, Phys.\ Rev. D 78, 046006 (2008)

\bibitem{sl15} P.A. Terzis and T. Christodoulakis, Class. Quantum Grav. 29,
235007 (29)

\bibitem{sl16} P.A. Terzis and T. Christodoulakis, Gen.\ Rel. Grav. 41, 469
(2009)

\bibitem{sl17} J.K. Erickson, D. H. Wesley, P.J. Steinhardt and\ N. Turok,
Phys.\ Rev. D 69, 063514 (2004)

\bibitem{steph2} M. MacCallum, Exact solutions in cosmology. In:
Hoenselaers, C., Dietz, W. (eds) Solutions of Einstein's Equations:
Techniques and Results. Lecture Notes in Physics, vol 205. Springer, Berlin,
Heidelberg, (1984)

\bibitem{Ratra} B.~Ratra and P.~J.~E.~Peebles, Phys. Rev. D 37 (1988)

\bibitem{Barrow} J.~D.~Barrow and P.~Saich, Class. Quant. Grav. 10 (1993)

\bibitem{q14} R.C. Caldwell, M. Kamionkowski and N.N. Weinberg, Phys. Rev.
Lett. 91, 071301 (2003)

\bibitem{q15} V. Faraoni, Phantom cosmology with general potentials, Class.
Quantum Grav. 22, 3235 (2005)

\bibitem{cop} E.J.\ Copeland, A.R. Liddle and D. Wands, Phys. Rev.\ D 57,
4686 (1998)

\bibitem{orl1} A. Belfiglio, O. Luongo and S. Mancini, Phys. Rev. D 107,
103512 (2023)

\bibitem{orl2} O. Luongo and T. Mengoni, Class. Quantum Grav. 41, 105006
(2024)

\bibitem{orl3} R. D'Agostino and O. Luongo, Phys. Lett. B 829, 137070 (2022)

\bibitem{orl4} O. Luongo, Gravitational metamaterials from optical
properties of spacetime media (2025) [arXiv:2504.09987]

\bibitem{sot} T.P. Sotiriou, Gravity and Scalar Fields. In: Papantonopoulos,
E. (eds) Modifications of Einstein's Theory of Gravity at Large Distances.
Lecture Notes in Physics, vol 892. Springer, Cham (2015)

\bibitem{gn1} O. Hrycyna and M. Szydlowski, JCAP 12, 016 (2013)

\bibitem{gn2} A. Coley and S. Hervik, Class. Quantum Grav. 22, 579 (2005)

\bibitem{gn3} R. De Arcia, I. Quiros, U. Nucamendi and T. Gonzales, Phys.\
Dark Univ. 40, 101183 (2023)

\bibitem{Russo} J.G. Russo, Phys. Lett. B 600, 185 (2004)

\bibitem{Basilakos} S. Basilakos, M. Tsamparlis and A. Paliathanasis, Phys.
Rev. D 83, 103512 2011

\bibitem{an1} C. Rubano and J. D. Barrow, Phys. Rev. D. 64, 127301 (2001)

\bibitem{an2} E. Piedipalumbo, P. Scudellaro, G. Esposito and C. Rubano,
Gen. Relativ. Gravit. 44 2611 (2012)

\bibitem{an3} A.\ Paliathanasis, M. Tsamparlis, S. Basilakos and J.D Barrow,
Phys.\ Rev. D 91, 123535 (2015)

\bibitem{an4} A.Yu. Kamenshchik, E.O. Pozdeeva, A. Tronconi, G. Venturi and
S.Yu. Vernov, Class. Quant. Grav. 33, 015004 (2016)

\bibitem{an4a} S.Yu Vernov, V.R. Ivanov and E.O. Pozdeeva, Phys.\ Part.
Nucl. 51, 744 (2020)

\bibitem{an5} L. A. Urena-Lopez, T. Matos, Phys. Rev. D 62, 081302 (2000)

\bibitem{an6} A.Yu. Kamenshchik, E.O. Pozdeeva, A.\ Tronconi,\ G. Venturi
and S.Yu. Vernov, Phys. Part. Nucl. 49, 1 (2018)

\bibitem{l1} P. Fre, A. Sagnotti and A.S. Sorin, Nucl. Phys. B 877, 1028
(2013)

\bibitem{l2} D. Fermi, M.\ Gengo, L. Pizzocchero, Nucl. Phys. B 957, 115095
(2020)

\bibitem{ll01} L.P. Eisenhart, Dynamical Trajectories and Geodesics, Annals.
Math. 30, 591 (1928)

\bibitem{ll02} C. Duval, G. Burdet, H.P. Kunzle and M. Perrin, Bargmann
structures and Newton-Cartan theory, Phys. Rev. D 1841, 31 (1985)

\bibitem{ll03} M. Cariglia and F.K. Alves, The Eisenhart lift: a didactical
introduction of modern geometrical concepts from Hamiltonian dynamics, Eur.
Phys. J. Plus 36, 025018 (2015)

\bibitem{lift1} A. Paliathanasis, Phys. Dark Univ. 44, 101466 (2024)

\bibitem{lift2} A. Paliathanasis, J. Geom. Phys. 206, 105338 (2024)

\bibitem{ll04} M. Cariglia and F.K. Alves, The Eisenhart lift: a didactical
introduction of modern geometrical concepts from Hamiltonian dynamics, Eur.
Phys. J. Plus 36, 025018 (2015)

\bibitem{ll05} G.K. Karananas, M. Michel,\ and J. Rubio, Phys. Lett. B 850,
138524 (2024)

\bibitem{ll06} M. Cariglia, A. Galajinsky, G.W. Gibbons and P.A. Horvathy,
Cosmological aspects of the Eisenhart--Duval lift, Eur. Phys. J. C 78, 314
(2018)

\bibitem{ani0} A. Paliathanasis, Int. J.\ Theor. Phys. 63, 303 (2024)

\bibitem{sin1} S. Cotsakis and P.G.L. Leach, J. Phys. A: Math. Gen. 27, 1625
(1994)

\bibitem{sin2} J. Demaret and C. Scheen, J. Math. Phys. A: Math. Gen. 29, 59
(1996)

\bibitem{sin3} A. Paliathanasis and P.G.L. Leach, Phys. Lett. A 380, 2815
(2016)

\bibitem{sin4} A. Paliathanasis, J.D. Barrow and P.G.L. Leach, Phys. Rev. D
94, 023525 (2016)

\bibitem{sin5} G. Leon, A.\ Paliathanasis P.G.L. Leach, Math.\ Meth. Appl.
Sci. 47, 6301 (2024)

\bibitem{sin6} S. Cotsakis, G. Kolionis and A. Tsokaros, Phys. Lett. B 721,
1 (2013)

\bibitem{sin7} S. Cotsakis, J. Demaret, Y. De Rop and L. Querella, Phys.
Rev. D 48, 4595 (1993)

\bibitem{sin8} W. Khyllep, A. Paliathanasis and J. Dutta, Phys. Rev.\ D 103,
103521 (2021)

\bibitem{Abl1} M.J. Ablowitz, A. Ramani and H. Segur, \ Lettere al Nuovo
Cimento 23, 333 (1978)

\bibitem{Abl2} M.J. Ablowitz, A. Ramani and H. Segur, J. Math. Phys. 21, 715
(1980)

\bibitem{Abl3} M.J. Ablowitz, A. Ramani and H. Segur, J. Math. Phys. 21,
1006 (1980)

\bibitem{buntis} A. Ramani, B. Grammaticos and T. Bountis, Physics Reports,
180 159 (1989)
\end{thebibliography}
\end{document}